\documentclass[a4paper,aps,pre,floatfix,twocolumn,nofootinbib,showpacs]{revtex4-1}

\usepackage{times}
\usepackage{verbatim}
\usepackage{bbold}
\usepackage{bbm}
\usepackage[pdftex]{graphicx}
\setkeys{Gin}{width=0.9\columnwidth}
\usepackage{latexsym,amsmath,verbatim}
\usepackage{color}
\usepackage{hyperref}
\usepackage{lipsum}
\usepackage{rotating}
\usepackage{multirow}
\usepackage[english]{babel}
\usepackage{microtype}
\usepackage{comment}
\usepackage[normalem]{ulem}
\usepackage{float}
\usepackage{subfigure}
\usepackage{colortbl}
\usepackage{enumitem}

\usepackage{array}
\newcolumntype{C}[1]{>{\centering}m{#1}}

\newcommand{\av}[1]{\langle #1 \rangle}

\begin{document}

\title{Flocking dynamics mediated by weighted social networks}

\author{Jaume Ojer}

\affiliation{Departament de F\'isica, Universitat Polit\`ecnica de
  Catalunya, Campus Nord, 08034 Barcelona, Spain}

\author{Romualdo Pastor-Satorras}

\affiliation{Departament de F\'isica, Universitat Polit\`ecnica de
  Catalunya, Campus Nord, 08034 Barcelona, Spain}

\date{\today}

\begin{abstract}
  We study the effects of animal social networks with a weighted pattern
  of interactions on the flocking transition exhibited by models of
  self-organized collective motion. Considering a model representing
  dynamics on a one-dimensional substrate, application of a
  heterogeneous mean-field theory provides a phase diagram as function
  of the heterogeneity of the network connections and the correlations
  between weights and degree. In this diagram we observe two phases, one
  corresponding to the presence of a transition and other to a
  transition suppressed in an always ordered system, already observed in
  the non-weighted case. Interestingly, a third phase, with no
  transition in an always disordered state, is also obtained. These
  predictions, numerically recovered in computer simulations, are also
  fulfilled for the more realistic Vicsek model, with movement in a
  two-dimensional space. Additionally, we observe at finite network
  sizes the presence of a maximum threshold for particular weight
  configurations, indicating that it is possible to tune weights to
  achieve a maximum resilience to noise effects. Simulations in real
  weighted animal social networks show that, in general, the presence of
  weights diminishes the value of the flocking threshold, thus
  increasing the fragility of the flocking state. The shift in the
  threshold is observed to depend on the heterogeneity of the weight
  pattern.
\end{abstract}

\maketitle

\section{Introduction}
\label{sec:introduction}

Many animal species are able to coordinate their behavior in base to the
individuals' interactions, resulting in the formation of self-organized
patterns of movement~\cite{camazine2003}. Such processes, broadly known
as collective motion~\cite{sumpter2010}, show stunning examples covering
widely separated time and length scales, ranging from the migration of
large mammals, the marching of huge groups of desert locusts or the
complex and coordinated maneuvering of flocks of birds and shoals of
fish, to the swimming and swarming of
bacteria~\cite{Sumpter2006,sumpter2010,vicsek2012,giardina2008,Cavagna2018}. The
field of collective motion has experienced recently an important boost
due to improvements in image acquisition and, especially, in tracking
technologies, capable to reconstruct the movement of many unmarked
individuals from digital
recordings~\cite{Romero-Ferrero:2019uo,walter2020trex}. However, most of
the scientific effort in the field has been devoted to the development
of models capable to explain and/or reproduce the features observed in
real groups of animals. Despite the different formulations of models of
collective motion, they are usually based in a set of moving
self-propelled particles (SPPs) implementing three main ingredients: (i)
avoiding collisions; (ii) trying to stay together; and (iii) trying to
align the velocity with that of the nearest
neighbors~\cite{19821081,Reynolds:1987:FHS:37402.37406,couzin2002}.

Most of these models consider metric interactions, where the neighbors
of the SPPs are defined in terms of Euclidean distance. It has been also
proposed that interactions might have in some cases a non-metric nature,
defined by a fixed number of closest neighbors, independently of their
relative distance~\cite{Ballerini.2008ufv,Ginelli2010}, and even by a
single closest neighbor in the forward
direction~\cite{Herbert-Read.2011}. These local, metric or non-metric,
rules, however, neglect the effect of possible \textit{social
  interactions} among the group members~\cite{croft2008exploring} and
that can induce individuals to try to follow with higher preference
other individuals that are closely connected socially with
them~\cite{Ling2019}.

The effect of social interactions in collective motion has been studied
in some detail in the context of the celebrated Vicsek
model~\cite{Vicsek1995}.  In this model, a set of SPPs move in a
two-dimensional space.  Dynamics evolves in discrete time, and is given
by the SPPs trying to align the direction of their velocity with the
average of a set of other SPPs in a local neighborhood. This alignment
is hindered by a source of noise of strength $\eta$, that represents
inherent difficulties in gathering the speed of the neighbors or in
implementing the resulting average. The interest of this minimal model
of collective motion resides in the fact that it exhibits an
order-disorder (\textit{flocking}) transition at a threshold value
$\eta_c$ of the noise intensity, separating an ordered phase at
$\eta \leq \eta_c$, in which particles move coherently in a randomly
chosen average direction, from a disordered phase at $\eta > \eta_c$, in
which SPPs behave as uncorrelated persistent random walkers. This model
has allowed to draw useful conclusions and analogies between the
collective motion of animals and the well-known features of
order-disorder phase transitions in classical statistical
mechanics~\cite{vicsek2012,Ginelli2016}, besides having been the subject
of many variations and modifications implementing possible realistic
features of animal
behavior~\cite{gregoire2004,Ginelli2010,Gao.2011,Clusella.2021}.

Social interactions are introduced in the Vicsek model in terms of a
complex network~\cite{Newman10}, in which nodes represent individuals
and connections among nodes the presence of social interactions between
pairs of individuals. The set of interacting neighbors of a SPP is thus
fixed and does not change in time, being given by the network adjacency
matrix $a_{ij}$ taking value $1$ when nodes $i$ and $j$ are socially
connected, and $0$ otherwise.  Several works have considered the effects
of different network topologies on the flocking transition experienced
by the Vicsek
model~\cite{Aldana2007,Pimentel2008,SEKUNDA2016,Bode2011a,bode2011}. An
interesting observation in this context are the effects that a
heterogeneous pattern of social interactions, observed in certain animal
social networks~\cite{Lusseau2003,Manno2008}, can have on the flocking
transition in the Vicsek model.  Ref.~\cite{Miguel2017} considered
heterogeneous complex topologies represented by networks with a degree
distribution $P(k)$, defined as the probability that a node is connected
to $k$ other nodes (i.e.~has degree $k$), with a power-law form,
$P(k) \sim k^{-\gamma}$~\cite{Barabasi:1999}. In this case, it was
observed that for a degree exponent $\gamma > 5/2$, a standard
transition is present, while for $\gamma < 5/2$, the transition is
suppressed, being the system in the ordered stated, in the thermodynamic
limit of infinite network size, for all physical values of the noise
strength $\eta$. The same particular role of the degree exponent was
recovered analytically in Ref.~\cite{Miguel.2019} using the scalar
version of the Vicsek model proposed by Czir\'ok, Barab\'asi and Vicsek,
the CBV model~\cite{Czirok1999}, in which velocity is a real number,
instead of a vector in a two-dimensional space. These results are
relevant for the understanding of the collective motion in social
animals, as they indicate that the flocking phase is more robust against
noise effects in the case of a highly heterogeneous pattern of social
contacts.

While the consideration of a networked pattern of contacts provides a
realistic setting for the influence of social relation in flocking
behavior, it still neglects the important fact that social networks have
an intrinsic weighted
nature~\cite{PhysRevE.70.056131,Barrat16032004,barthelemy2005}, which
reflects the obvious fact that not all social connections have the same
strength, in the sense, for example, that a close friend can exert a
stronger influence than a casual acquaintance. Such a weight pattern has
been shown to have important effects on dynamical processes running on
top of
them~\cite{0256-307X-22-2-068,karsai:036116,Deijfen201157,Baronchelli.2011}
and even to be relevant for the efficient transfer of information
between social animals~\cite{Rosenthal2015}.

In this paper we explore the effects of a weighted pattern of social
contacts on collective motion by considering the flocking transition of
the vectorial Vicsek model and the related scalar CBV model, when placed
on top of a weighted network. We focus in particular in the case of
heterogeneous networks, empirically observed in certain animal social
networks~\cite{Lusseau2003,Manno2008}, with a degree distribution of the
form $P(k) \sim k^{-\gamma}$. The weighted structure is defined by a
pattern of weights $w_{ij}$, a set of real positive numbers representing
the strength of the social tie between individuals $i$ and $j$. For the
case of the CBV model, and considering a weight pattern depending on the
degree of the connected nodes of the form
$w_{ij} = (k_i k_j)^\alpha a_{ij}$, as observed in many real
systems~\cite{Barrat16032004}, we develop a theoretical approach based
in the heterogeneous mean-field theory
(HMF)~\cite{pv01a,10.5555/1521587,dorogovtsev07:_critic_phenom,Baronchelli.2011,Pastor-Satorras:2014aa}. The
theory provides a phase diagram for the behavior of the flocking
transition threshold $\eta_c$ in the thermodynamic limit as a function
of the degree exponent $\gamma$ of the degree distribution and the
weight exponent $\alpha$. This phase diagram recovers the results
observed in the unweighted case, namely a phase with a true transition
at a finite $\eta_c$ value and a phase where the transition is absent
and the system is always ordered. Surprisingly, however, a new phase
emerges, in which the system is \textit{always} disordered, in the
infinite size limit, for any value of $\eta$ however small. In this
phase, the systems becomes extremely sensitive to the effects of noise,
with a flocking phase that can be destroyed even for small values of
$\eta.$ Additionally, in the case of networks of finite size, we observe
that the predicted threshold in a given network has a maximum value for
a particular weight exponent, which indicates that we can engineer the
resilience of the system to external disorder (i.e.~maximize $\eta_c$)
for a particularly chosen weight structure.

These theoretical predictions are confirmed by means of computer
simulations of the CBV model. In the case of the Vicsek model, while
lacking an explicit theoretical formulation, we observe numerically that
the results for the CBV model can be extrapolated by just taking into
account the proper physical limits of the noise parameter $\eta$ in each
model. We finally consider the Vicsek model on real animal social
weighted networks. We observe that, in real weighted networks, the
effect of a weight structure consists in decreasing the transition
threshold with respect to the binary, non-weighted network. This
indicates that the actual weight structure makes animal social networks
more fragile to external noise. While no theory is again available for
real networks, we empirically observe that the shift in the threshold
observed in weighted networks can be related to the degree of
heterogeneity of the weight pattern.

\section{Models of flocking dynamics in weighted networks}
\label{sec:vect-scal-flock}

In this Section we describe the implementation on weighted networks of
two models of collective motion, the classical Vicsek
model~\cite{Vicsek1995}, in which particles move on a two-dimensional
space with a vectorial velocity, and the CBV model~\cite{Czirok1999},
which represents individuals moving on a line and characterized by a
scalar velocity.

\subsection{Vectorial Vicsek model}
\label{sec:vicsek-model}

The Vicsek model is defined in terms of a set of $N$ SPPs moving in a
two-dimensional space, characterized by a position $\mathbf{r}_i(t)$ and
a velocity $\mathbf{v}_i(t)$ at time $t$. Dynamics is defined in
discrete time and velocities are assumed to have a constant modulus,
$\lvert \mathbf{v}_i(t) \rvert = v_0$, and are thus determined by the
angle $\theta_i(t)$ they form with the $x$ axis, taking the form
\begin{equation}
  \label{eq:1}
  \mathbf{v}_i(t) = v_0 \cos \theta_i(t)\; \mathbf{\hat{i}} +  v_0 \sin
  \theta_i(t) \;\mathbf{\hat{j}} .
\end{equation}
In the original Vicsek model~\cite{Vicsek1995}, each SPP $i$ tends to
align its velocity parallel to the average velocity $\mathbf{V}_i$ of a
set of SPPs in a local neighborhood inside a circle of radius $R$
centered at $i$. In the case of an unweighted (binary) network,
interactions are constant and defined by the nearest neighbors connected
to a node. Thus, in terms of the adjacency matrix, the dynamics of
velocities is defined by the synchronous update rule~\cite{Miguel2017}
\begin{equation}
  \label{eq:2}
   \theta_i(t+1) = \Theta\left[ \mathbf{v}_i(t) + \sum_{j=1}^N a_{ij}
     \mathbf{v}_j(t)\right] + \eta \xi_i(t),
\end{equation}
where the function $\Theta[\mathbf{V}]$ returns the angle described by a
vector $\mathbf{V}$, $\xi_i(t)$ is random noise uniformly distributed in
the interval $[-\pi, \pi]$, and $\eta \in [0,1]$ is a parameter
measuring the strength of the external noise. We notice that, with this
definition, the noise strength has a maximum value $\eta=1$, compatible
with a complete randomization of the information provided by the average
velocity of the nearest neighbors.

In the case of weighted networks~\cite{Barrat16032004}, a real positive
number $w_{ij}$ is assigned to the edge connecting nodes $i$ and $j$,
representing the strength of the social interaction between individuals
$i$ and $j$. Here we will consider the case of undirected weighted
networks, in which $w_{ij} = w_{ji}$, i.e.~the influence of node $i$
over node $j$ is exactly the same as that exerted over $i$ by $j$. When
placed on top of a weighted network, we define the Vicsek update rule by
\begin{equation}
  \label{eq:3}
  \theta_i(t+1) = \Theta\left[ \mathbf{v}_i(t) + \frac{k_i \sum_{j=1}^N
      w_{ij} \mathbf{v}_j(t)}{\sum_{r=1}^N w_{ir}} \right] + \eta
  \xi_i(t).
\end{equation}
With this rule, we consider that the average velocity of the neighbors
of agent $i$ is computed giving a normalized weight
$w_{ij} / [ \sum_r w_{ir} / k_i ]$ to each neighbor $j$, where the
normalization factor has been chosen as the average weight of all nodes
adjacent to $i$, in such a way that the limit to a constant value
$w_{ij} = w_0$ recovers the dynamics in unweighted networks,
Eq.~\eqref{eq:2}.

In many real weighted networks, the weight of the edge connecting nodes
$i$ and $j$ is found to be a function of the product of the degrees of
the connected nodes~\cite{Barrat16032004},
\begin{equation}
  \label{eq:4}
  w_{ij} = w_0 (k_i k_j)^\alpha a_{ij},
\end{equation}
$\alpha$ being an exponent characterizing the correlation between weight
and degrees.  In this case, the interaction rule takes the simplified
form
\begin{equation}
  \label{eq:5}
  \theta_i(t+1) = \Theta\left[ \mathbf{v}_i(t) + \frac{k_i \sum_{j=1}^N
      k_j^\alpha a_{ij} \mathbf{v}_j(t)}{\sum_{r=1}^N k_r^\alpha a_{ir}}
  \right] + \eta \xi_i(t).
\end{equation}
The order parameter for the Vicsek model in networks is defined as in
the spatial version, namely
\begin{equation}
  \label{eq:6}
  \phi(\eta) = \lim_{T\to\infty} \frac{1}{v_0 T N} \sum_{t' = t_m}^{t_m
    +T} \left|\sum_{i=1}^N \mathbf{v}_i(t') \right|,
\end{equation}
where $t_m$ is a sufficiently large thermalization time.

\subsection{Scalar CBV model}
\label{sec:scalar-cbv-model}

The scalar CBV model~\cite{Czirok1999} is defined by a set of $N$ SPPs
on a one-dimensional substrate, in which particles move with velocity
$u_i(t)$. Each SPP $i$ updates its velocity considering the local
average velocity $U_i$ of other agents in a neighborhood
$\left[ x_i - \Delta, x_i + \Delta \right]$ surrounding it. This average
velocity is modulated by a function $G(U)$, that restricts the
individual velocities to remain close to $+1$ or $-1$, in order to avoid
diverging trajectories. Individual velocities are finally updated by
this modulated local average velocity with the addition of a noise
term. For a binary network, the update rule can be defined
as~\cite{Miguel.2019}
\begin{equation}
  \label{eq:7}
  u_i(t+1) = G \left[ \frac{\sum_j a_{ij} u_j(t)}{k_i} \right] + \eta
  \xi_i(t),
\end{equation}
where $\xi_i$ is a uniform random number in the interval $[-1/2, 1/2]$
and $\eta \in [0,\infty)$ gauges the strength of the external noise. For
simplicity, the modulating function $G(U)$ is chosen to be the sign
function, taking value $G(U) = +1$ when $U \geq 0$ and $G(U) = -1$
otherwise~\cite{Miguel.2019}. We notice that, in this prescription, we
do not consider the interaction of the velocity of a node with itself.

In the case of a weighted network, the update rule can be easily
extended from the Vicsek model, taking the form
\begin{equation}
  \label{eq:8}
  u_i(t+1) = G \left[\sum_j w_{ij} u_j(t)  \right] + \eta \xi_i(t),
\end{equation}
where we have discarded irrelevant factors due to the nature of the sign
function $G(U)$. When the weights have the topological structure given
by Eq.~\eqref{eq:4}, the update rule can be further simplified as
\begin{equation}
  \label{eq:9}
   u_i(t+1) = G \left[\sum_j k_j^\alpha a_{ij} u_j(t)  \right] + \eta
   \xi_i(t).
\end{equation}
The order parameter is defined in this case
as~\cite{Czirok1999,Miguel.2019}
\begin{equation}
  \label{eq:10}
  \phi(\eta) = \lim_{T\to\infty} \frac{1}{TN} \sum_{t' = t_m}^{t_m +T}
  \left | \sum_{i=1}^N u_i(t') \right |.
\end{equation}

\section{Heterogeneous mean-field theory for the CBV model in weighted
  networks}
\label{sec:heter-mean-field}

The CBV model in weighted networks can be tackled numerically applying
the HMF approximation developed in Ref.~\cite{Miguel.2019} (see also
\cite{Chen2015}). We start by rewriting the update dynamics in terms of
the dual velocities $u^*_i$ as
\begin{eqnarray}
  u_i^*(t+1) &=& G\left[\sum_j k_j^\alpha a_{ij}
                 u_j(t)\right], \label{eq:11} \\
  u_i(t+1) &=&  u_i^*(t+1) + \eta \xi_i, \label{eq:12}
\end{eqnarray}
from where it is easy to see that the dual velocities fulfill
\begin{equation}
  \label{eq:13}
  u_i^*(t+1) = G\left [\sum_j k_j^\alpha a_{ij}  u_j^*(t) +  \eta
      \sum_j k_j^\alpha a_{ij} \xi_j  \right] .
\end{equation}
Due to the sign function $G$, the dual velocities are spin variables,
$u^*_i = \{ -1, 1\}$, a fact that greatly simplifies the subsequent
analysis. To solve the dynamics of the dual velocities, we apply a HMF
approach inspired in Refs.~\cite{Castellano2006,Chen2015,Miguel.2019},
assuming that all dynamical properties of nodes are a function of their
degree alone, in such a way that nodes with the same degree $k$,
defining a degree class, share the same dynamical properties. We define
$\rho_k(t)$ as the probability that a randomly chosen node of degree $k$
is in state $+1$ at time $t$, and $\psi_k(t)$ as the probability that a
randomly chosen node of degree $k$ will flip to the state $+1$ at time
$t$. These two quantities are related by the rate equation
\begin{eqnarray}
  \dot{\rho}_k(t) &=& - \rho_k(t) [1 - \psi_k(t)] + [1 - \rho_k(t)]
  \psi_k(t) \nonumber \\
  &=& - \rho_k(t) + \psi_k(t) \label{eq:14},
\end{eqnarray}
which, in the steady state $\dot{\rho}_k(t) = 0$, leads to
\begin{equation}
  \label{eq:15}
  \rho_k = \psi_k.
\end{equation}

Consider now the dynamics of Eq.~\eqref{eq:13}, where we drop the star
superindex to ease notation.  The function $\psi_k$ can be computed
considering a node $i$ of degree $k$ and computing its probability to
flip to a spin value $+1$. From Eq.~(\ref{eq:13}), this probability is
equal to the probability that the argument $R$ inside the sign function
$G$ is positive. This argument can be written as the sum of two
contributions, $R = R_u(k) + R_\xi(k)$, with
\begin{equation}
  \label{eq:28}
  R_u(k) = \sum_j a_{ij} k_j^\alpha u_j, \; \; R_\xi(k) = \eta \sum_j a_{ij}
  k_j^\alpha \xi_j .
\end{equation}
Starting with the second term, it corresponds to a random variable equal
to the sum of $k$ random variables $ \eta k_j^\alpha \xi_j$ of mean zero
and variance $\sigma^2_j = k_j^{2\alpha} \sigma^2_0$, where
$\sigma^2_0 = \eta^2 / 12$ is the variance of the original noise term
$\xi_j$. In the HMF approximation, the neighbors $j$ are chosen at
random in an uncorrelated network~\cite{alexei} with probability
$P_n(k_j) = \frac{k_j P(k_j)}{\av{k}}$, depending only on their
degree. By the central limit theorem, we can thus see that $R_\xi(k)$ is
a Gaussian random variable of mean zero and variance
\begin{equation}
  \label{eq:17}
  \sigma^2 = k \sum_{k_j} \frac{k_j P(k_j)}{\av{k}} \sigma^2_j =
  k\sigma_0^2 \frac{\av{k^{1+2\alpha}}}{\av{k}}.
\end{equation}
The factor $R_u(k)$ is more difficult to estimate probabilistically, so
we will only consider its average value. $R_u(k)$ is given by the sum of
the contributions $k_j^\alpha u_j$ for the nearest neighbors $j$ of node
$i$. Considering that the variable $u_j$ in a node of degree $k'$ takes
value $+1$ with probability $\rho_{k'}$, the average value of $R_u(k)$
is given by
\begin{eqnarray}
  \bar{R_u}(k) &=& k \sum_{k'} \frac{k' P(k')}{\av{k}} k'^{\alpha}
  \left[ (+1) \rho_{k'} + (-1)(1-\rho_{k'}) \right] \nonumber \\
  &=& k \sum_{k'} \frac{k'^{1+ \alpha} P(k')}{\av{k}}
  \left[ 2 \rho_{k'} - 1 \right] =
  k \frac{\av{k^{1+\alpha}}}{\av{k}} q,  \label{eq:18}
\end{eqnarray}
where the factor
\begin{equation}
  \label{eq:19}
  q = \sum_{k} \frac{k^{1+\alpha} P(k)}{\av{k^{1+\alpha}}}
  \left [2\rho_k - 1 \right]
\end{equation}
plays the role of an effective order parameter, with value $q=0$ in the
disordered state, where $\rho_k= 1/2$, and $q \neq 0$ in the ordered
state $\rho_k\neq 1/2$.

The probability $\psi_k$ is thus equal to the probability that
$R = R_\xi(k) + \bar{R_u}(k)$ is larger than zero. Since $R_\xi(k)$ is a
Gaussian variable of zero mean and variance Eq.~(\ref{eq:17}), we can
write
\begin{eqnarray}
  \psi_k &=& \int_{-\bar{R_u}(k)}^\infty \frac{1}{\sqrt{2 \pi \sigma^2}}
  e^{-r^2 / (2 \sigma^2)}\; dr \nonumber \\
  &=& \frac{1}{2} + \frac{1}{2} \mathrm{erf}
  \left( \frac{\bar{R_u}(k)}{\sqrt{2} \sigma} \right) \nonumber \\
  &=& \frac{1}{2} + \frac{1}{2} \mathrm{erf} \left( \sqrt{k}
  \frac{\av{k^{1+\alpha}}}{[\av{k} \av{k^{1+2\alpha}}]^{1/2} }
  \frac{q}{\sigma_0 \sqrt{2}} \right),  \label{eq:20}
\end{eqnarray}
where $\mathrm{erf}(z)$ is the error function~\cite{abramovitz}. In the
steady state $\psi_k = \rho_k$, so we can compute $q$ self-consistently
from Eq.~(\ref{eq:20}) as
\begin{eqnarray}
  q &=& \sum_{k} \frac{k^{1+\alpha} P(k)}{\av{k^{1+\alpha}}}
  \left [2 \psi_k - 1 \right]  \equiv F(q) \nonumber \\
  &=& \sum_{k} \frac{k^{1+\alpha} P(k)}{\av{k^{1+\alpha}}} \mathrm{erf}
  \left( \sqrt{\frac{k}{2}}
  \frac{\av{k^{1+\alpha}}}{[\av{k} \av{k^{1+2\alpha}}]^{1/2}}
  \frac{q}{\sigma_0} \right).
      \label{eq:21}
\end{eqnarray}
The equation $q = F(q)$ has a nonzero solution, corresponding to the
onset of the ordered state, when the first derivative of $F(q)$
evaluated at $q=0$ is larger than one, that is, when
\begin{eqnarray}
  F'(0) &=& \sum_{k} \frac{k^{1+\alpha} P(k)}{\av{k^{1+\alpha}}}
  \frac{2}{\sqrt{\pi}} \sqrt{\frac{k}{2}}
  \frac{\av{k^{1+\alpha}}}{[\av{k} \av{k^{1+2\alpha}}]^{1/2}}
  \frac{1}{\sigma_0} \nonumber \\
  &=& \sqrt{\frac{2}{\pi}} \frac{1}{\sigma_0}
  \frac{\av{k^{3/2 + \alpha}}}{[\av{k} \av{k^{1+2\alpha}}]^{1/2}} > 1.
  \label{eq:22}
\end{eqnarray}
From here, a threshold condition appears,
\begin{equation}
  \label{eq:23}
  \sigma_0 < \sqrt{\frac{2}{\pi}}
  \frac{\av{k^{3/2 + \alpha}}}{[\av{k} \av{k^{1+2\alpha}}]^{1/2} },
\end{equation}
that, in terms of the noise intensity $\eta = \sqrt{12} \sigma_0$,
allows to define the noise threshold
\begin{equation}
  \label{eq:24}
  \eta_c = \sqrt{\frac{24}{\pi}}
  \frac{\av{k^{3/2 + \alpha}}}{[\av{k}\av{k^{1+2\alpha}}]^{1/2}},
\end{equation}
such that an ordered state is present for $\eta < \eta_c$, and a
disordered one for $\eta > \eta_c$. We notice here the presence of an
erroneous factor $2$ in Eq.~(29) of Ref.~\cite{Miguel.2019}, which
renders it equal to our general prediction Eq.~\eqref{eq:24} for
$\alpha=0$ in the limit of large threshold\footnote{In
  Ref.~\cite{Miguel.2019} the term corresponding to the factor $R_u(k)$
  was treated probabilistically and not in average value. This explains
  that the result here and there only coincide in the limit of large
  threshold.}.

\begin{figure} [t]
  \includegraphics[width=0.95\columnwidth]{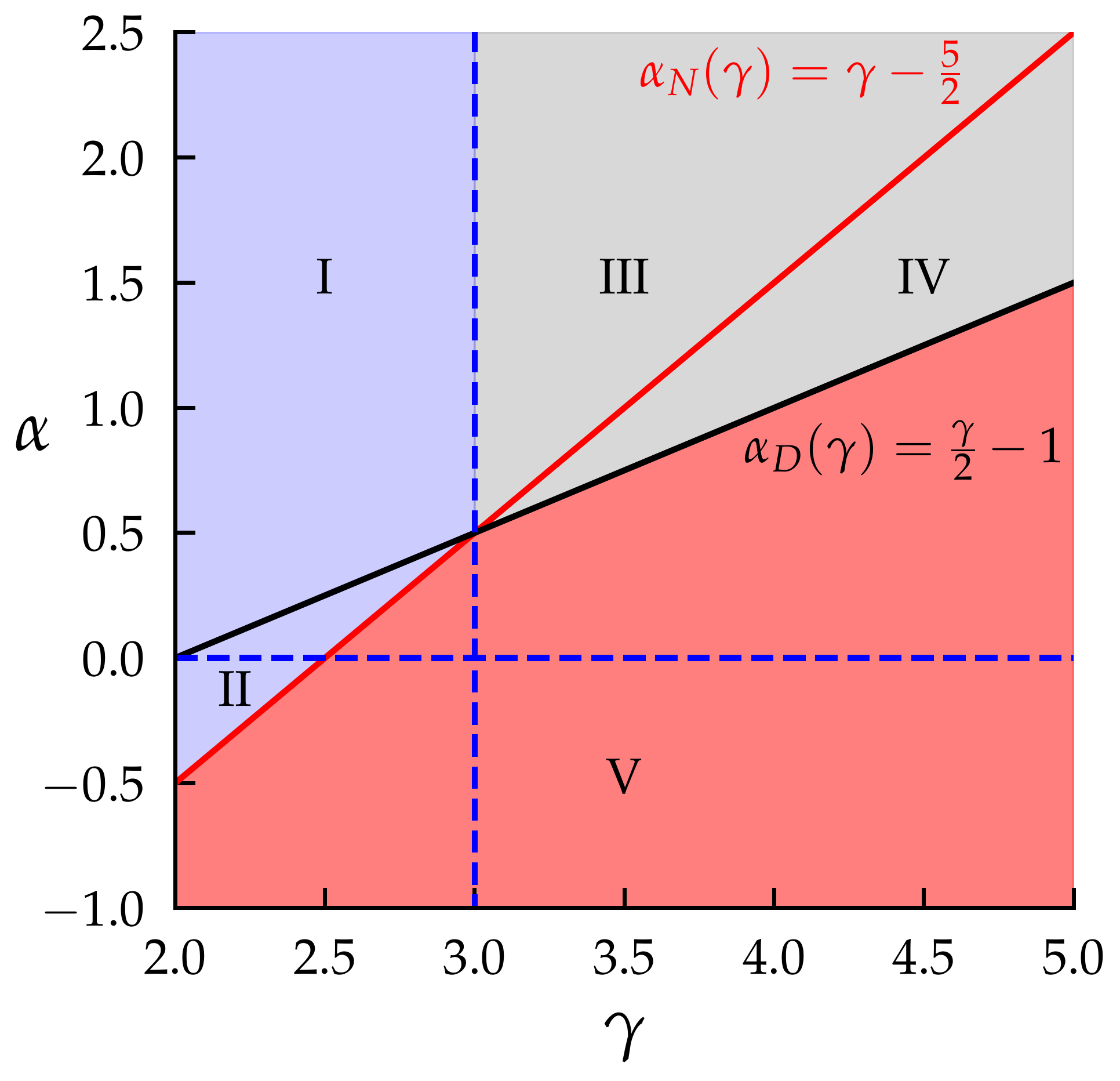}
  \caption{Phase diagram of the CBV model on weighted networks. Red and
    black lines mark respectively the functions $\alpha_N(\gamma)$ and
    $\alpha_D(\gamma)$, defined in Eq.~\eqref{eq:27}. The vertical
    dashed line indicates the value $\gamma=3$. The horizontal dashed
    line indicates the value $\alpha=0$, corresponding to an unweighted
    network. In regions I and II (shaded in blue), the threshold
    diverges in the thermodynamic limit, $\eta_c \to \infty$; in regions
    III and IV (shaded in gray), the threshold converges to zero,
    $\eta_c \to 0$; in region V (shaded in red), the threshold converges
    to a constant, $\eta_c \to \mathrm{const}$.}
  \label{fig:phase_diagram}
\end{figure}

In the case of interest of scale-free networks with a degree
distribution $P(k) \sim k^{-\gamma}$, the value of the noise threshold
of the CBV model in weighted networks depends on ratios of moments that
can lead to peculiar behavior in the thermodynamic limit depending on
$\alpha$ and $\gamma$. Assuming $\gamma > 2$, in order to ensure a
sparse network with constant average degree $\av{k}$, the value of the
threshold depends on the moment ratio
$\eta_c \sim \av{k^{3/2 + \alpha}} / \av{k^{1+2\alpha}}^{1/2}$.
Defining the functions
\begin{eqnarray}
  \label{eq:27}
  \alpha_N(\gamma) = \gamma - \frac{5}{2}, \quad \alpha_D(\gamma) =
  \frac{\gamma}{2} -1,
\end{eqnarray}
we can see that, in a network with a maximum degree
$k_c$~\cite{dorogovtsev07:_critic_phenom,Boguna09}, the numerator of
Eq.~\eqref{eq:24} diverges in the thermodynamic limit $k_c \to \infty$
as $\av{k^{3/2 + \alpha}} \sim k_c^{\alpha - \gamma + \frac{5}{2}}$ for
$\alpha > \alpha_N(\gamma)$, while it goes to a constant for
$\alpha < \alpha_N(\gamma)$. On the other hand, the denominator diverges
as $\av{k^{1+2\alpha}}^{1/2} \sim k_c^{\alpha - \frac{\gamma}{2} + 1}$
for $\alpha > \alpha_D(\gamma)$, going instead to a constant for
$\alpha < \alpha_D(\gamma)$. This leads to different scaling behaviors
of the noise threshold in the thermodynamic limit that are summarized in
the phase diagram portrayed in Fig.~\ref{fig:phase_diagram}. In regions
I and III, both numerator and denominator diverge, leading to
$\eta_c \sim k_c^{(3 - \gamma)/2}$. Thus, in region I, with
$\gamma < 3$, the noise threshold diverges, while it converges to zero
in region III. In region II, numerator diverges and denominator
converges, and so the threshold diverges. In region IV, numerator and
denominator exchange behavior, and thus the threshold converges to
zero. Finally, in region V, both denominator and numerator converge, and
the threshold converges to a constant.

The scaling of the threshold with the network size $N$ can be recovered
if we consider, for uncorrelated networks, that $k_c \sim N^{1/2}$ for
$\gamma < 3$ and that $k_c \sim N^{1/(\gamma - 1)}$ for
$\gamma > 3$~\cite{mariancutofss}. We therefore obtain, in the limit of
large $N$ and in the different regions:
\begin{enumerate}[align=left,leftmargin=35pt,label=Region \Roman*:]
\item $\eta_c \sim N^{(3 - \gamma)/4} \to \infty$;
\item $\eta_c \sim N^{[2(\alpha - \gamma) + 5]/4} \to \infty$;
\item $\eta_c \sim N^{-(\gamma - 3)/[2(\gamma - 1)]} \to 0$;
\item $\eta_c \sim N^{-[2(\alpha + 1) - \gamma]/[2(\gamma - 1)]} \to 0$;
\item $\eta_c \to \mathrm{const}$.
\end{enumerate}

This analytical solution recovers the main result in
Refs.~\cite{Miguel2017,Miguel.2019} regarding the presence of a phase in
which a true transition is present, characterized by a finite threshold,
separated from another region in which the threshold tends to its
maximum physical value in the thermodynamic limit, indicating that the
transition is absent and, therefore, the system is always ordered for
any value of $\eta$. These regions now depend on the values of $\alpha$
for $\gamma <3$. The most noticeable feature of this solution, however,
is the emergence of a new phase, regions III and IV, in which a set of
values of $\alpha$ for $\gamma >3$ lead to a null threshold in the
thermodynamic limit. This case corresponds again to the absence of
transition, but now in a system that is always in the disordered state,
no matter how small the noise strength might be.

\section{Numerical results in synthetic weighted networks}
\label{sec:numer-results-synth}

In order to check the analytical predictions obtained in the previous
Section, as well as to obtain a more precise rendering of the effects of
a weighted topology on the ordering dynamics of the CBV and Vicsek
model, in this Section we consider numerical simulations of both models
on synthetic heterogeneous networks with a scale-free degree
distribution given by a power-law form, $P(k) \sim k^{-\gamma}$. In
particular, we generate networks using the uncorrelated configuration
model (UCM)~\cite{Catanzaro05} with a minimum degree
$k_\mathrm{min} = 3$ and a maximum degree
$k_c = \min (N^{1/2}, N^{1/(\gamma-1)})$, in order to avoid degree
correlations and maximum degree
fluctuations~\cite{mariancutofss,Boguna09}. On these networks, we impose
a weight on each edge given by Eq.~\eqref{eq:4}. The parameters of the
network models are thus the degree exponent $\gamma$ and the weight
exponent $\alpha$. In our simulations, we compute statistical quantities
allowing for a thermalization time $t_m = 50,000$ and averaging over
$T = 250,000$ time steps for the CBV model. For the Vicsek model, we
choose $t_m = 10,000$ and $T = 50,000$.

\subsection{CBV model}
\label{sec:cbv-model-real}

\begin{figure}[t]
  \includegraphics[width=0.9\columnwidth]{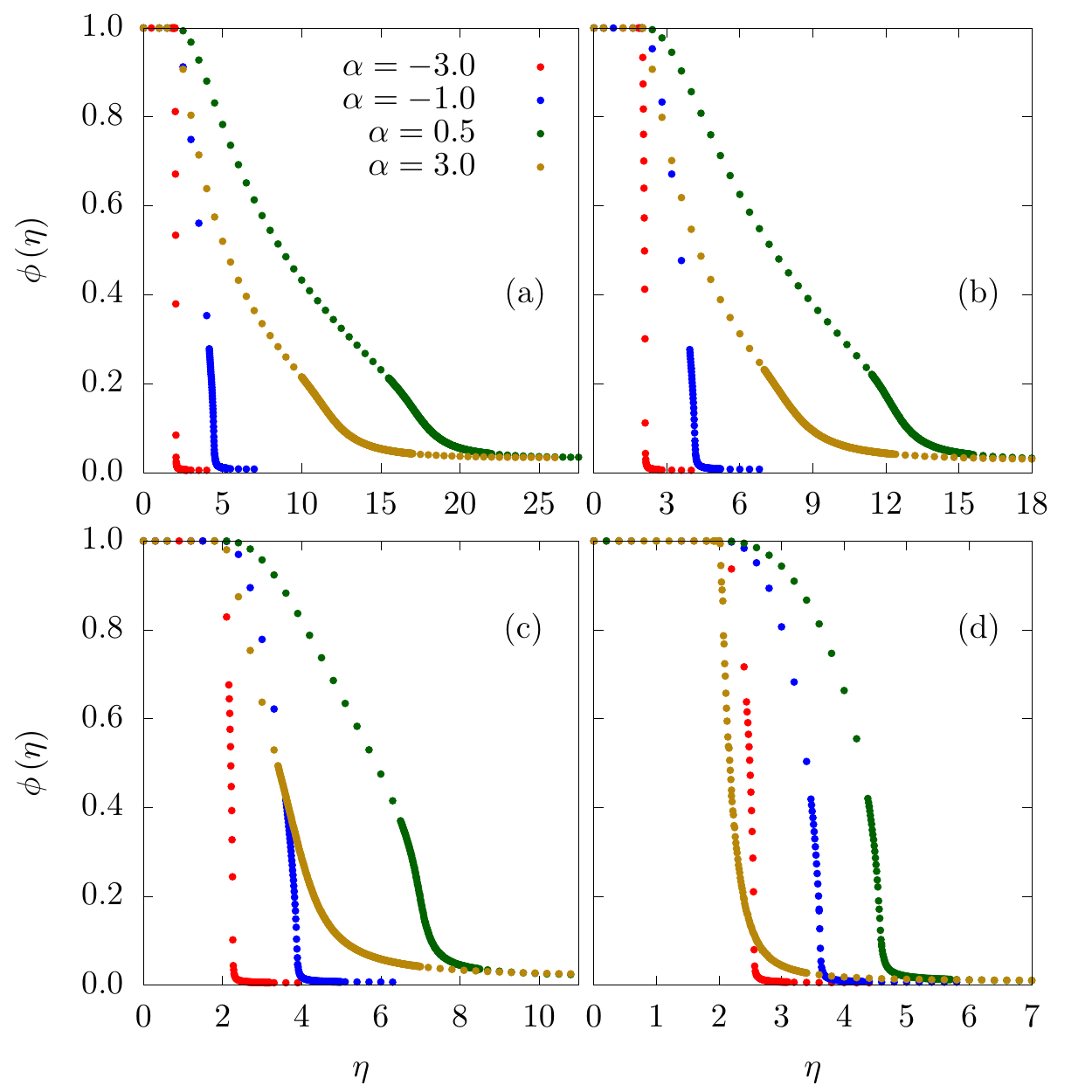}
  \caption{Order parameter $\phi(\eta)$ as a function of $\eta$ in the
    CBV model on weighted UCM networks with different degree ($\gamma$)
    and weight ($\alpha$) exponents. Panels correspond to different
    values of the degree exponent: (a) $\gamma=2.10$, (b) $\gamma=2.35$,
    (c) $\gamma=2.75$, (d) $\gamma=3.50$. Network size $N=10^5$.}
  \label{fig:CBV_orderparameter}
\end{figure}

\begin{figure} [t]
  \includegraphics[width=0.9\columnwidth]{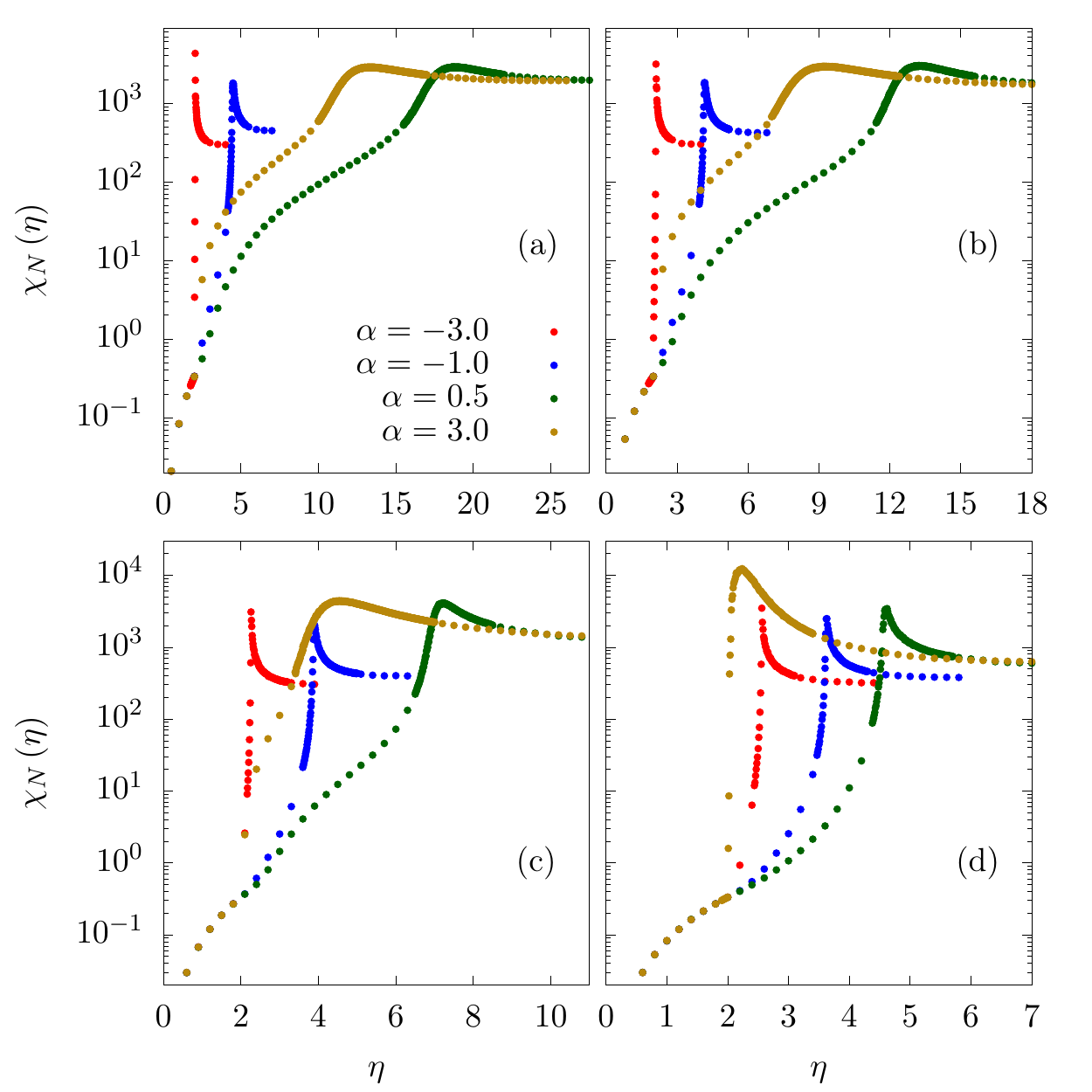}
  \caption{Dynamic susceptibility $\chi_N(\eta)$ as a function of $\eta$
    in the CBV model on weighted UCM networks with different degree
    ($\gamma$) and weight ($\alpha$) exponents. Panels correspond to
    different values of the degree exponent: (a) $\gamma=2.10$, (b)
    $\gamma=2.35$, (c) $\gamma=2.75$, (d) $\gamma=3.50$. Network size
    $N=10^5$.}
  \label{fig:CBV_susceptibility}
\end{figure}

In the first place, we check the predictions of the HMF theory developed
in Sec.~\ref{sec:heter-mean-field} for the CBV model on weighted
networks. In Fig.~\ref{fig:CBV_orderparameter} we show the order
parameter $\phi(\eta)$ as a function of the noise intensity $\eta$
computed in networks of different degree and weight exponents. As we can
see from this Figure, the order parameter is compatible with the
presence of a threshold, that depends in a complex way on both exponents
$\gamma$ and $\alpha$.  In order to determine this threshold noise in
simulations on necessarily finite systems, we consider the dynamic
susceptibility, defined as~\cite{Ferreira2012,Castellano2016}
\begin{equation}
  \label{eq:26}
  \chi_N(\eta) = N \frac{\av{\phi^2} - \av{\phi}^2 }{\av{\phi}}.
\end{equation}
The effective critical point $\eta_c(N)$ in a network of size $N$ is
given by the value of the noise at the maximum of the susceptibility
$\chi_N(\eta)$~\cite{Ferreira2012,Castellano2016,Miguel2017,Miguel.2019}. In
Fig.~\ref{fig:CBV_susceptibility} we plot the shape of the dynamic
susceptibility computed from a sample of values of $\gamma$ and
$\alpha$. As we can see, a clear peak is observed in all plots, that
allows to define the effective threshold as a function of the network
size, $\eta_c(N)$. At this peak, the maximum value of the dynamic
susceptibility, $\chi^{\mathrm{peak}}(N) \equiv \chi_N(\eta_c(N))$, is
expected to show a power-law increase with the network size, given
by~\cite{Ferreira2012,Miguel.2019}
\begin{equation}
  \label{eq:16}
  \chi^{\mathrm{peak}}(N) \sim N^\delta,
\end{equation}
where $\delta$ is a characteristic exponent.

\begin{figure} [t]
  \includegraphics[width=0.9\columnwidth]{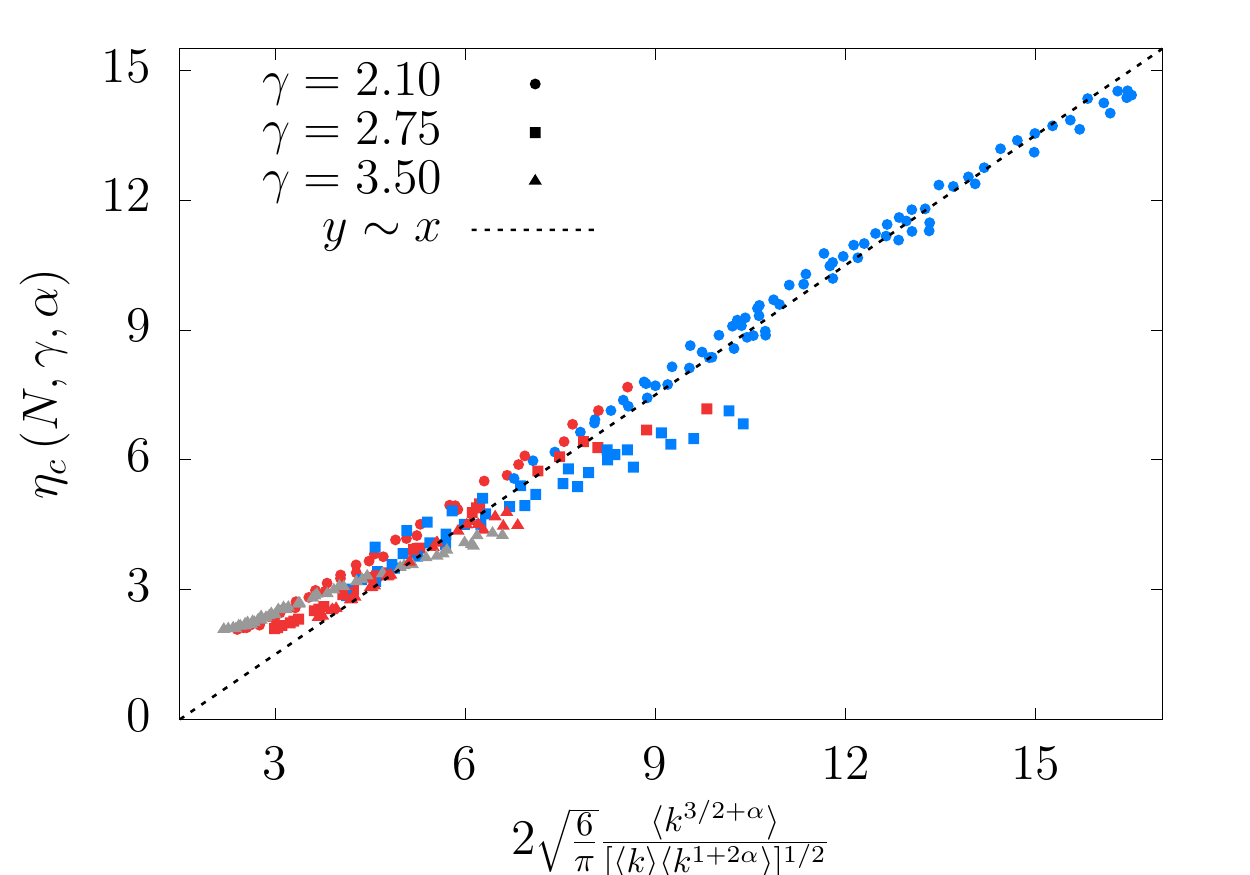}
  \caption{Effective critical point $\eta_c(N, \gamma, \alpha)$ as a
    function of the theoretical prediction, Eq.~(\ref{eq:24}), in the
    CBV model on weighted UCM networks with different degree ($\gamma$)
    and weight ($\alpha$) exponents and different network sizes $N$. The
    values of $\alpha$ and $N$ considered range in the intervals
    $[-3, 4]$ and $[10^3, 10^5]$, respectively. The color of the symbols
    denote their position in the phase diagram
    Fig.~\ref{fig:phase_diagram}: blue for regions I and II
    ($\eta_c \to \infty$); gray for regions III and IV ($\eta_c \to 0$);
    red for region V ($\eta_c \to \mathrm{const}$).}
  \label{fig:comparison}
\end{figure}

In Fig.~\ref{fig:comparison} we compare the effective threshold
$\eta_c(N, \gamma, \alpha)$ in the CBV model, estimated by the peak of
the dynamic susceptibility, with the theoretical HMF prediction in
Eq.~\eqref{eq:24}, for different values of the degree exponent $\gamma$,
weight exponent $\alpha$ and network size $N$. As we can see,
disregarding a common vertical intercept, the theoretical prediction
provides a very good approximation to the numerical values observed in
simulations. The fit is particularly good for region I (blue circles)
and region V (red symbols), where the threshold is expected to diverge or
converge to a constant, respectively, in the thermodynamic limit.

\begin{figure} [t]
  \includegraphics[width=0.9\columnwidth]{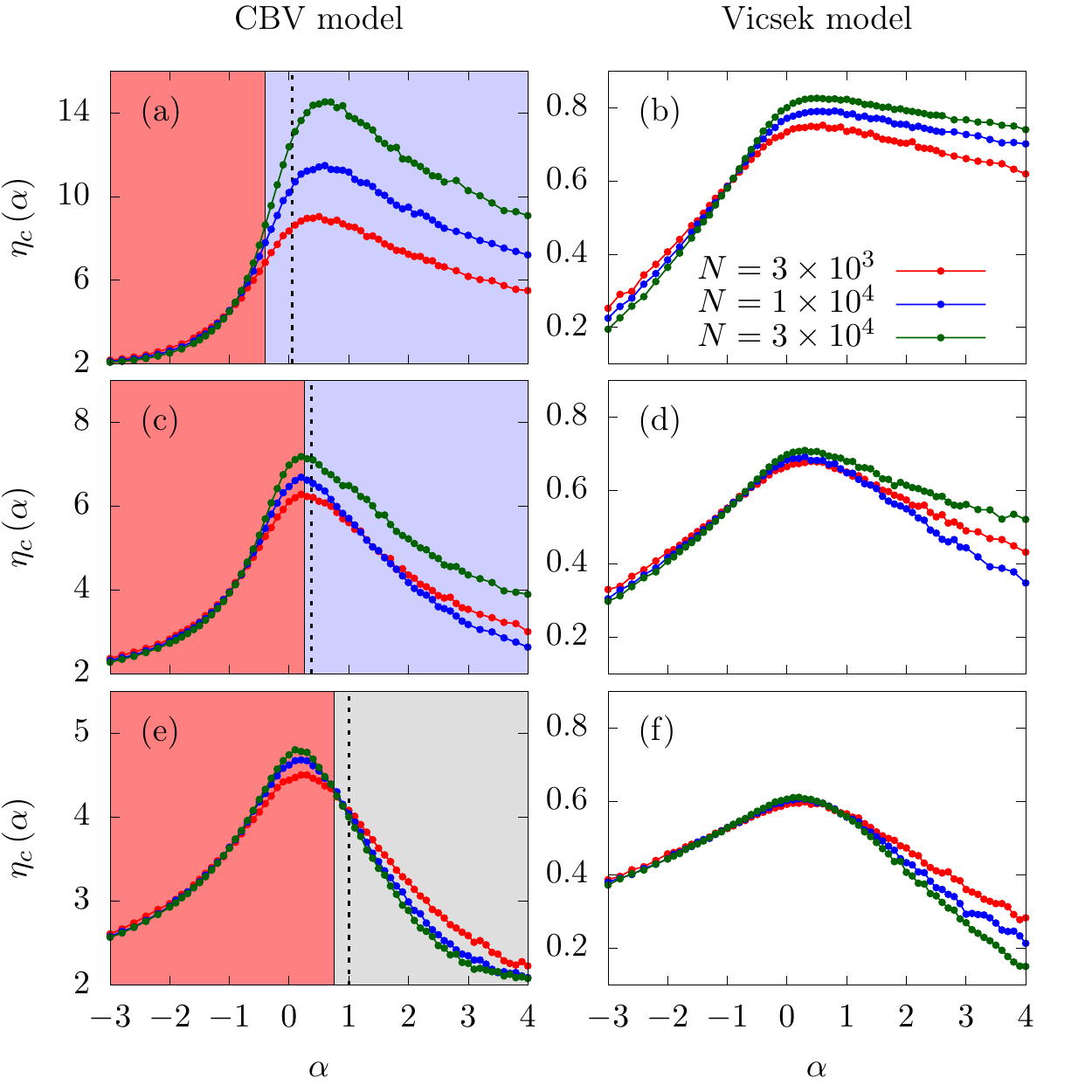}
  \caption{Effective critical point $\eta_c(\alpha)$ as a function of
    the weight exponent $\alpha$ on UCM networks of different size
    $N$. Left panels show results corresponding to the CBV model. In
    this case, we have color-marked the region in the phase diagram
    Fig.~\ref{fig:phase_diagram} corresponding to the expected scaling
    of the threshold: red (region V) for $\eta_c \to \mathrm{const}$,
    blue (regions I and II) for $\eta_c \to \infty$; gray (regions III
    and IV) for $\eta_c \to 0$.  Vertical dashed lines mark the
    transition from region I to II, and from III to IV. Right panels
    correspond to simulations of the Vicsek model. Values of gamma are:
    (a) and (b), $\gamma = 2.10$; (c) and (d), $\gamma = 2.75$; (e) and
    (f), $\gamma = 3.50$.}
  \label{fig:simulations}
\end{figure}

We now verify the scaling behavior of the threshold in the different
regions represented in the phase diagram in
Fig.~\ref{fig:phase_diagram}. To do so, in Fig.~\ref{fig:simulations}
(left panels) we plot the effective threshold $\eta_c(\alpha)$, measured
as the peak of the dynamic susceptibility, as a function of the weight
exponent $\alpha$ in networks of different degree exponent $\gamma$, for
different network sizes $N$. From this Figure, we can see that the
numerical thresholds indeed recover the scaling form resulting from the
HMF analysis. For $\gamma =2.1$, the values of $\alpha$ in region V lead
to an effective threshold converging to a constant as we increase the
network size. For values of $\alpha$ in regions I and II, on the other
hand, the threshold is observed to diverge for increasing $N$. For
$\gamma = 2.75$, small values of $\alpha$ in region V again lead to a
constant threshold. However, the situation for larger values of $\alpha$
is more complex, due to the fact that this value of $\gamma$ is quite
close to the singular case of $\gamma = 3$ for which
$\alpha_N(\gamma) = \alpha_D(\gamma)$ and all regions coalesce. One
would need much larger network sizes to observe the theoretical
prediction for the thermodynamic limit. For $\gamma=3.5$ we recover in
region V (small $\alpha$) a converging threshold. For large $\alpha$ in
regions III and IV, however, we observe the interesting feature of a
threshold that tends to zero when increasing the network size. Again,
finite size effects affect the behavior in the boundary between regions
V and IV.

Another important feature that can be observed from
Fig.~\ref{fig:simulations} (left panel) is the presence of a maximum in
the threshold $\eta_c(\alpha)$ as a function of $\alpha$. This maximum
reflects the fact that the resilience of the system to the effects of
noise is maximal for a weight exponent $\alpha_\mathrm{max}$, depending
in principle on the degree exponent. This observation can be recovered
analytically by setting equal to zero the derivative of
Eq.~(\ref{eq:24}) with respect to $\alpha$ and solving the ensuing
equation, that leads to $\alpha_\mathrm{max}^\mathrm{HMF} = 1/2$ for any
$\gamma$. The threshold at this maximum depends on the second moment of
the degree distribution, $\eta_c^\mathrm{max} \sim \av{k^2}^{1/2}$,
diverging in the thermodynamic limit for $\gamma <3$ (region I) and
converging to a constant for $\gamma > 3$ (region V). In numerical
simulations, Fig.~\ref{fig:simulations} (left panel), the maximum
$\alpha_\mathrm{max}$ is clearly present, but it seems to depend on the
degree exponent and to slightly change with the network size. In order
to check this, in Table~\ref{tab:alpha_max} we summarize the variation
of the maximum $\alpha_\mathrm{max}(\gamma)$ estimated numerically as we
increase $N$ and depending on the heterogeneity of the network. These
numerical results show that HMF analysis provides a very good prediction
for small values of the degree exponent, with
$\alpha_\mathrm{max} \simeq 0.5$ for $\gamma=2.1$. For larger values of
$\gamma$ we obtain a more complex dependence. Thus, for large $N$ and
$\gamma=2.35$ we have $\alpha_\mathrm{max} \simeq 0.7$, while for
$\gamma \geq 2.75$ we observe $\alpha_\mathrm{max} \simeq 0.1$.

\begin{table}[b]
\begin{tabular}{>{\centering}p{0.17\columnwidth}>{\centering}p{0.18\columnwidth}>{\centering}p{0.19\columnwidth}>{\centering}p{0.19\columnwidth}>{\centering\arraybackslash}p{0.18\columnwidth}}
\noalign{\global\arrayrulewidth=0.2mm}
\hline
\multirow{2}{*}{$N$} & \multicolumn{4}{c}{$\alpha_\mathrm{max}(\gamma)$} \\ \noalign{\global\arrayrulewidth=0.1mm} \cline{2-5}
& $\gamma = 2.10$ & $\gamma = 2.35$ & $\gamma = 2.75$ & $\gamma = 3.50$ \\
\noalign{\global\arrayrulewidth=0.2mm}
\hline
$3 \times 10^3$ & $0.5$ & $0.6$ & $0.2$ & $0.3$ \\
$1 \times 10^4$ & $0.6$ & $0.5$ & $0.2$ & $0.2$ \\
$3 \times 10^4$ & $0.6$ & $0.6$ & $0.2$ & $0.1$ \\
$1 \times 10^5$ & $0.6$ & $0.7$ & $0.1$ & $0.1$ \\
$3 \times 10^5$ & $0.5$ & $0.8$ & $0.1$ & $0.1$ \\
\arrayrulecolor{black}\hline
\end{tabular}
\caption{Numerical estimation of the weight exponent
  $\alpha_\mathrm{max}(\gamma)$ for which the effective threshold is
  maximum,
  $\eta_c^\mathrm{max} \equiv \eta_c(\alpha_\mathrm{max}(\gamma))$, in
  the CBV model on UCM weighted networks of different degree exponent
  $\gamma$ and size $N$. The error in the estimation of the maxima is
  $\Delta \alpha_\mathrm{max} = 0.1$ in all cases.}
\label{tab:alpha_max}
\end{table}

\begin{figure} [t]
  \includegraphics[width=0.9\columnwidth]{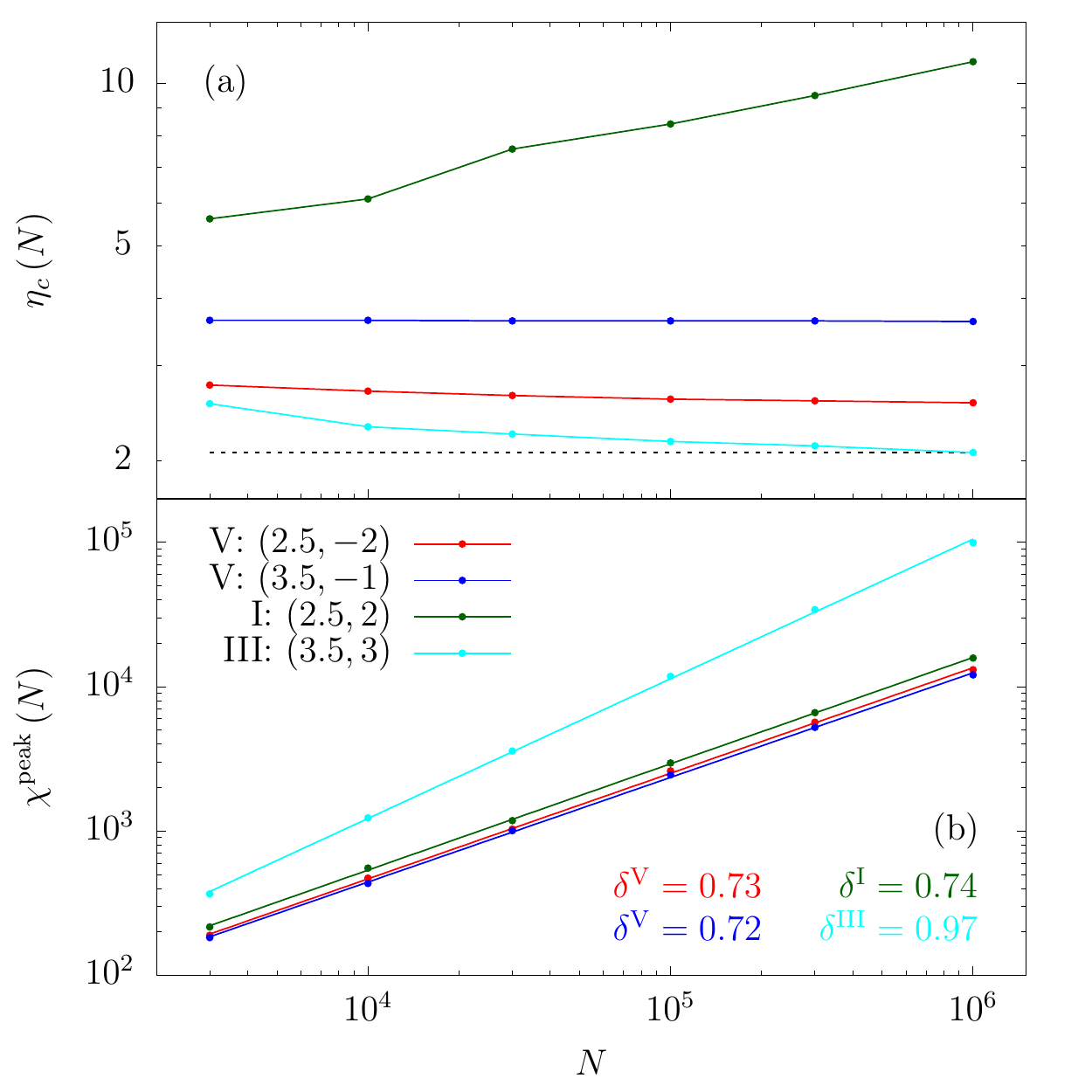}
  \caption{(a) Effective critical point $\eta_c(N)$ as a function of the
    network size $N$ for different pairs of values ($\gamma, \alpha$)
    belonging to regions I, III and V. The dashed horizontal line serves
    to highlight the slow decay to zero of the threshold observed in
    region III (b) Maximum of the dynamic susceptibility at the peak,
    $\chi^{\mathrm{peak}}(N)$, as a function of the network size $N$ for
    different pairs of values ($\gamma, \alpha$). The exponents $\delta$
    quoted of the different regions are obtained by means of a linear
    regression in log-log scale to the form
    $\chi^\mathrm{peak} \sim N^\delta$.  Results correspond to the CBV
    model on UCM weighted networks.}
  \label{fig:CBV_finitesize}
\end{figure}

Finally, in Fig.~\ref{fig:CBV_finitesize} we study in more detail the
finite size scaling of the CBV model as a function of network size $N$
for different points ($\gamma, \alpha$) belonging to regions I, III and
V. We do not consider regions II and IV since it is difficult to select
points sufficiently away from the boundaries $\gamma=3$,
$\alpha_N(\gamma)$ and $\alpha_D(\gamma)$, without choosing extremely
large values of $\gamma$ and $\alpha$. In
Fig.~\ref{fig:CBV_finitesize}(a) we plot the resulting evolution of the
effective threshold as a function of $N$. The points corresponding to
region V show a very clear plateau, indicative that the constant
threshold predicted by HMF is quickly reached for moderate network
sizes. On the other hand, for the points in regions I and III, the
threshold shows an increasing and decreasing trend, respectively. The
increase of threshold with $N$ in region I is very clear,
while the decrease in region III is weaker. This fact can be understood
at the HMF level from the scaling of the threshold as a function of $N$
given in Sec.~\ref{sec:heter-mean-field}. In region I, we have
$\eta_c^\mathrm{I}(N) \sim N^{(3 - \gamma)/4}$, which for our sample
point ($2.5, 2$) leads to $\eta_c^\mathrm{I}(N) \sim N^{1/8}$. In region
III, instead,
$\eta_c^\mathrm{III}(N) \sim N^{-(\gamma - 3)/[2(\gamma - 1)]}$, that
for the sample point ($3.5, 3$) yields
$\eta_c^\mathrm{III}(N) \sim N^{-1/10}$, that is, a smaller exponent
than that expected in region I. We notice however that, despite this
argument is qualitatively correct, our numerical simulation do not
recover the exponents predicted by the theory.

In Fig.~\ref{fig:CBV_finitesize}(b) we study the behavior of the maximum
value of the dynamic susceptibility at its peak,
$\chi^{\mathrm{peak}}(N)$, as a function of $N$. In accordance with the
theoretical expectation, Eq.~\eqref{eq:16}, we observe that the peak of
the susceptibility increases with network size as a power-law,
$\chi^{\mathrm{peak}}(N) \sim N^\delta$. The characteristic growth
exponent $\delta$ seems to be constant and the same in regions I and V,
$\delta \simeq 0.73$, and instead it is quite larger in region III,
$\delta \simeq 0.97$.

\subsection{Vicsek model}
\label{sec:vicsek-model-1}

We have also performed numerical simulations of the vectorial Vicsek
model in weighted UCM networks. In this case, we do not have an explicit
analytical solution. We can however extrapolate the results of the CBV
model pursuing the analogy made in the case of binary networks with
$\alpha=0$~\cite{Miguel.2019}. Since the noise intensity is bounded by
the maximum value $1$ in the Vicsek model, we can interpret the
different regions of the phase diagram in the CBV model directly, just
considering that regions I and II, where the CBV model exhibits a
diverging threshold, correspond in the Vicsek case to a threshold that
saturates to the maximum value $1$ in the thermodynamic limit.

In Fig.~\ref{fig:simulations} (right panel) we present the evolution of
the effective threshold $\eta_c(\alpha)$ in the Vicsek model as a
function of the weight exponent $\alpha$ for UCM networks of different
degree exponent and size. A comparison with the corresponding plots for
the CBV case presented in the left panel shows that both models exhibit
the same trend in the behavior of the threshold for different values of
$\alpha$. This indicates that both models have an analogous phase
diagram, as long as a diverging threshold in regions I and II in the CBV
model is interpreted as a threshold converging to $1$ in the Vicsek
model. This observation provides further confirmation of the fact that
the dimensionality of the order parameter does not play a relevant role
in the characterization of the behavior of critical transitions in
networks~\cite{Miguel2017,dorogovtsev07:_critic_phenom}.

\begin{figure} [b]
  \includegraphics[width=0.9\columnwidth]{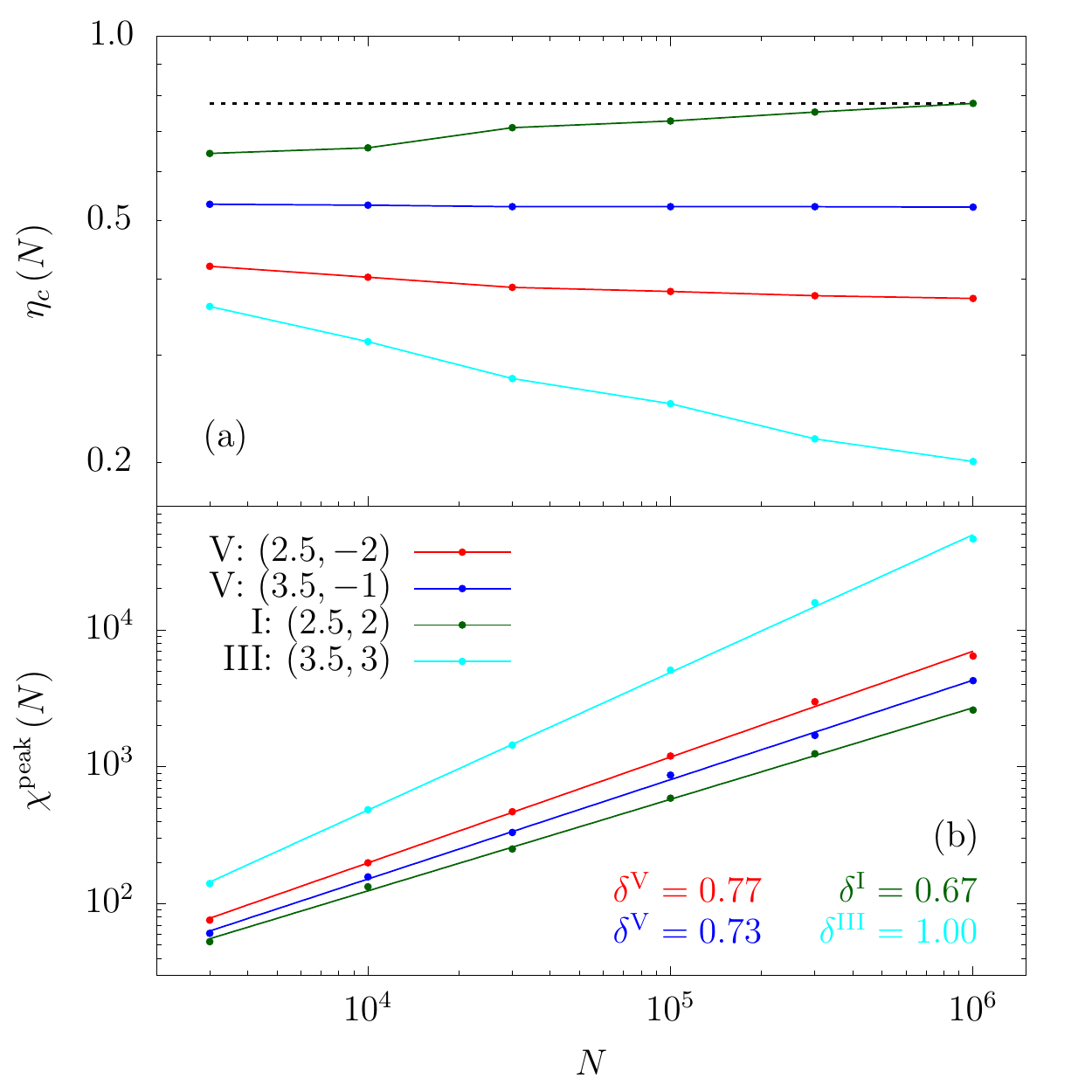}
  \caption{(a) Effective critical point $\eta_c(N)$ as a function of the
    network size $N$ for different pairs of values ($\gamma, \alpha$)
    belonging to regions I, III and V. The dashed horizontal line serves
    to highlight the slow convergence to $1$ of the threshold observed
    in region I (b) Maximum of the dynamic susceptibility at the peak,
    $\chi^{\mathrm{peak}}(N)$, as a function of the network size $N$ for
    different pairs of values ($\gamma, \alpha$). The exponents $\delta$
    quoted of the different regions are obtained by means of a linear
    regression in log-log scale to the form
    $\chi^\mathrm{peak} \sim N^\delta$.  Results correspond to the
    Vicsek model on UCM weighted networks.}
  \label{fig:Vicsek_finitesize}
\end{figure}

\begin{table}[b]
  \begin{ruledtabular}
  \begin{tabular}{|l|ccc|ccc|}
    Network & $N$ & $\av{w}$ & $\chi_w$ & $\eta_c^\mathrm{w}$ & $\eta_c^\mathrm{b}$ & $\Delta \eta$ \\
    \hline
    \texttt{Bison}~\cite{lott_american_1991} & $26$ & $2.76$ & $0.54$ & $0.834$ & $0.854$ & $0.023$ \\
    \texttt{Cattle}~\cite{hass_social_1991} & $28$ & $2.30$ & $0.64$ & $0.802$ & $0.843$ & $0.049$ \\
    \texttt{Sheep}~\cite{schein_social_1955} & $28$ & $2.66$ & $0.65$ & $0.829$ & $0.855$ & $0.030$ \\
    \texttt{Hyenas}~\cite{holekamp_society_2012} & $36$ & $0.08$ & $0.55$ & $0.866$ & $0.890$ & $0.027$ \\
    \texttt{Bats}~\cite{silvis_roosting_2014} & $43$ & $30.62$ & $1.17$ & $0.860$ & $0.885$ & $0.028$ \\
    \texttt{Sparrows}~\cite{shizuka_across-year_2014} & $46$ & $2.92$ & $1.09$ & $0.748$ & $0.828$ & $0.097$ \\
    \texttt{Dolphins 1}~\cite{hunt_assortative_2019} & $50$ & $0.33$ & $1.16$ & $0.730$ & $0.882$ & $0.172$ \\
    \texttt{Lizards}~\cite{bull_social_2012} & $60$ & $0.02$ & $8.10$ & $0.316$ & $0.724$ & $0.564$ \\
    \texttt{Squirrels}~\cite{smith_split_2018} & $61$ & $0.14$ & $1.28$ & $0.827$ & $0.872$ & $0.052$ \\
    \texttt{Thornbills}~\cite{farine_social_2013} & $62$ & $2.49$ & $0.72$ & $0.868$ & $0.884$ & $0.018$ \\
    \texttt{Macaques 1}~\cite{takahata1991} & $62$ & $2.06$ & $0.44$ & $0.851$ & $0.887$ & $0.041$ \\
    \texttt{Macaques 2}~\cite{balasubramaniam_social_2018} & $78$ & $2.53$ & $0.73$ & $0.862$ & $0.896$ & $0.038$ \\
    \texttt{Songbirds}~\cite{adelman_feeder_2015} & $110$ & $0.02$ & $3.03$ & $0.524$ & $0.802$ & $0.347$ \\
    \texttt{Ants}~\cite{mersch_tracking_2013} & $113$ & $7.06$ & $1.25$ & $0.858$ & $0.905$ & $0.052$ \\
    \texttt{Wildbirds}~\cite{firth_experimental_2015} & $149$ & $0.07$ & $1.27$ & $0.804$ & $0.852$ & $0.056$ \\
    \texttt{Dolphins 2}~\cite{gazda_importance_nodate} & $151$ & $1.21$ & $0.19$ & $0.813$ & $0.819$ & $0.007$ \\
    \texttt{Crickets}~\cite{fisher_comparing_2016} & $161$ & $2.78$ & $1.07$ & $0.481$ & $0.574$ & $0.162$ \\
    \texttt{Voles}~\cite{davis_spatial_2015} & $255$ & $2.19$ & $0.81$ & $0.421$ & $0.518$ & $0.187$ \\
    \texttt{Mice}~\cite{lopes_infection-induced_2016} & $280$ & $4.55$ & $3.93$ & $0.181$ & $0.313$ & $0.422$ \\
    \texttt{Sealions}~\cite{schakner_social_nodate} & $1007$ & $0.03$ & $0.63$ & $0.926$ & $0.939$ & $0.014$ \\
  \end{tabular}
\end{ruledtabular}
\caption{Topological properties of the real weighted networks
  analyzed. Network size $N$; average weight $\av{w}$; normalized
  variance of the weights $\chi_w = \av{w^2}/\av{w}^2 - 1$; effective
  threshold of the weighted network version $\eta_c^\mathrm{w}$;
  effective threshold of the binary network version $\eta_c^\mathrm{b}$;
  relative difference of the threshold in the weighted over binary
  networks $\Delta \eta = 1 - \eta_c^\mathrm{w} / \eta_c^\mathrm{b}$.}
  \label{tab:realnetsproperties}
\end{table}

In Fig.~\ref{fig:Vicsek_finitesize}(a) we show the analogous scaling
with network size of the effective threshold of the Vicsek model in the
same representative points of the different regions of the phase
diagram. As we can see, in full agreement with the observations for the
CBV model, region V leads to thresholds saturating to a constant value,
region III is characterized by a threshold decreasing with network size,
while region I shows an increasing threshold, necessarily saturating to
the maximum value $\eta=1$. Interestingly, the rate of decrease of the
threshold in region III is substantially larger in the Vicsek model than
in the CBV model, whereas the opposite happens for the rate of growth in
region I, being faster in the CBV model. This is due to the fact that,
in the Vicsek model, the threshold converges to a maximum value, while
in the CBV model it grows without limit.

Finally, in Fig.~\ref{fig:Vicsek_finitesize}(b) we present the growth of
the maximum of the susceptibility at its peak as a function of the
network size, for the different pairs of values ($\gamma, \alpha$). A
linear regression in logarithmic scale shows the expected power-law
dependence $\chi^\mathrm{peak}(N) \sim N^\delta$. In contrast with the
CBV model, in the Vicsek case the exponent $\delta$ seems to depend on
$\alpha$ and $\gamma$ simultaneously.

\section{Numerical results in real weighted networks}
\label{sec:numer-results-real}

\begin{figure*}[t]
  \includegraphics[width=0.8\textwidth]{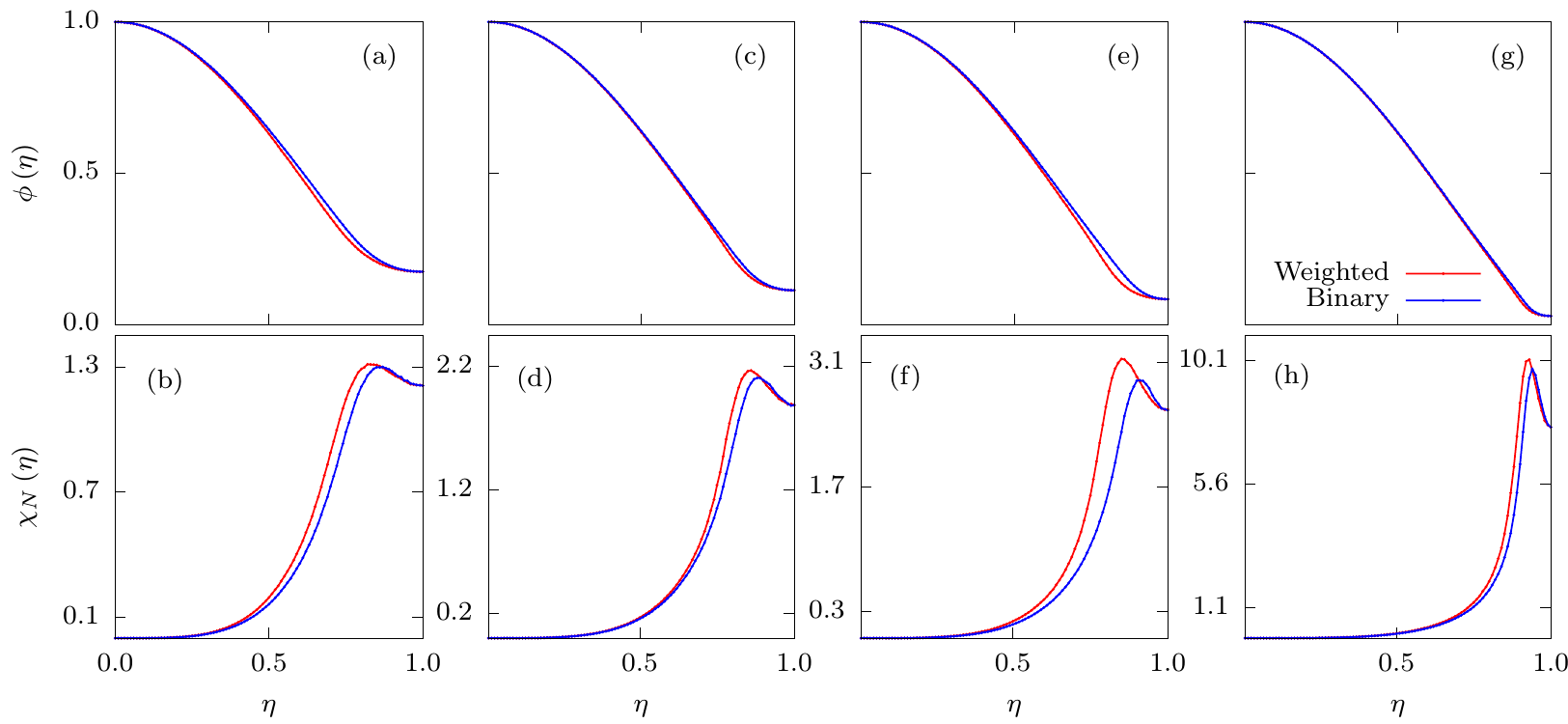}
  \caption{Order parameter $\phi(\eta)$ (top row) and dynamic
    susceptibility $\chi_N(\eta)$ (bottom row) as a function of $\eta$
    in the Vicsek model on a sample of four different real animal social
    weighted networks: (a) and (b) \texttt{Bison}, (c) and (d)
    \texttt{Macaques 1}, (e) and (f) \texttt{Ants}, (g) and (h)
    \texttt{Sealions}. A comparison between weighted and binary
    structures is shown.}
  \label{fig:realnetsdynamics}
\end{figure*}

In order to ascertain the effects of a real weighted topology on
flocking dynamics, we have studied the behavior of the Vicsek model on
several empirical animal social weighted
networks~\cite{anima_networks}. As weighted substrates, we consider
twenty networks reflecting dominance relationships, behavioral
activities, sexual interactions and mating associations in different
species (see Table~\ref{tab:realnetsproperties} for references to the
network's details).  Since some of these networks are directed in
nature, in our analysis we have worked with their undirected version, in
which weights have been symmetrized, defining
$w^s_{ij} = (w_{ij} + w_{ji}) / 2$. We have also disregarded nodes of
zero degree and edges of zero weight. For the simulations, we have set
$t_m = 50,000$ and $T = 500,000$. In Table~\ref{tab:realnetsproperties}
we present a summary of the topological properties of the animal
weighted networks considered.

The HMF theory developed in Sec.~\ref{sec:heter-mean-field} cannot be
directly applied to real networks, since those are usually
correlated~\cite{Newman10} and the relation between the weight of an
edge and the degrees at its endpoints is only approximately fulfilled
for large networks~\cite{Barrat16032004} and difficult to asses in small
ones. For this reason, in order to characterize the effects of weights
in our empirical networks, we have compared the behavior in the actual
weighted network with that of its binary projection, constructed by
assigning to all edges a constant weight $w_0$, arbitrarily fixed to
$1$.

In Fig.~\ref{fig:realnetsdynamics} we present a plot of the order
parameter $\phi(\eta)$ (top row) and the dynamic susceptibility
$\chi_N(\eta)$ (bottom row) as a function of the noise intensity $\eta$
for a sample of four real networks, comparing the results for the
weighted and binary simulation procedures. As we can see, the effect of
the weights in all four cases is to decrease the shape of the order
parameter of the weighted networks with respect to the binary version,
effectively reducing the degree of order for large values of $\eta$. At
the same time, we can see that the peak of the dynamic susceptibility is
shifted to the left in the weighted case, indicating that the effective
threshold in the weighted network, $\eta_c^\mathrm{w}$, is smaller than
in its binary counterpart, $\eta_c^\mathrm{b}$. This effect is confirmed
in the whole set of $20$ networks considered, as shown by the relative
threshold difference,
$\Delta \eta = 1 - \eta_c^\mathrm{w} / \eta_c^\mathrm{b}$, being always
positive, see Table~\ref{tab:realnetsproperties}.

\begin{figure} [h]
  \includegraphics[width=0.85\columnwidth]{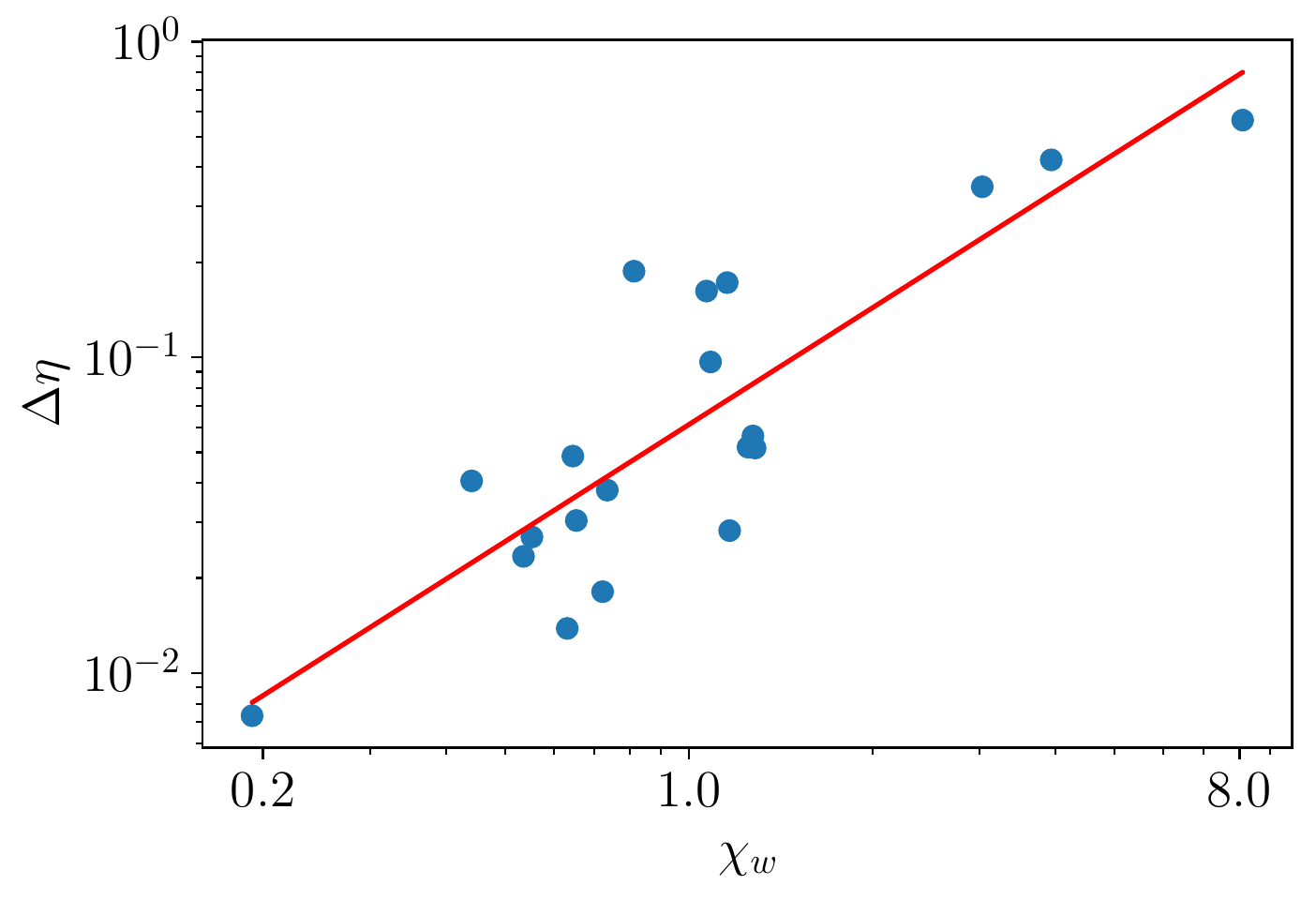}
  \caption{Relative difference of the weighted and binary thresholds
    $\Delta \eta$ as a function of the normalized variance of the
    weights $\chi_w$. A power law relation is denoted by the straight
    line, $\Delta \eta \sim \chi_w^{1.2}$.}
  \label{fig:noise_vs_heterogeneity}
\end{figure}

While we do not have an analytical insight about the dependence of the
threshold on the topological weighted substrate of the network, an
examination of Table~\ref{tab:realnetsproperties} shows that the
threshold in the weighted networks is correlated with the weight
heterogeneity, as measured by the normalized variance
$\chi_w = \av{w^2} / \av{w}^2 - 1$. Indeed, a closer inspection
indicates a stronger correlation between the relative threshold
difference $\Delta \eta$ and the variance of weights, which seems to
be related by a power-law form $\Delta \eta \sim \chi_w^{a}$, with an
exponent approximately equal to $a=1.2$, see
Fig.~\ref{fig:noise_vs_heterogeneity}. This exponent is obtained via
linear regression in double logarithmic scale, with a significant
Pearson regression coefficient $r = 0.85$.

\section{Conclusions}
\label{sec:conclusions}

Social ties can play an important role in shaping the interactions
between animals ruling their collective
behavior~\cite{Ling2019}. Indeed, it has been recently shown that a
heterogeneous pattern of social interactions, represented in terms of a
complex network, can alter the properties of the flocking transition in
simple models of collective behavior~\cite{Miguel2017,Miguel.2019}. Here
we have presented an extension of previous studies of collective motion
mediated by social interactions by considering the weighted nature of
social contacts, in which the network topology is enriched by adding a
real variable $w_{ij}$ between the connected nodes $i$ and $j$,
representing the intensity of the social contact between this pair of
individuals. In this context, we have focused on the effects of a
weighted topology on the threshold marking the position of the flocking
transition in the classic Vicsek model of collective motion, and in a
related model (CBV) which has been shown to behave similarly to the
Vicsek model in networks~\cite{Miguel.2019}. In the case of a weighted
structure depending on the degree of nodes, of the form
$w_{ij} = (k_i k_j)^\alpha a_{ij}$, the CBV model can be solved within a
heterogeneous mean-field approximation, obtaining an expression of the
threshold as a function of the exponent $\alpha$ and the exponent
$\gamma$ of the degree distribution, assumed to have a power-law form
$P(k) \sim k^{-\gamma}$. The solution provides a phase diagram in the
plane $(\gamma, \alpha)$, in which a phase corresponds to a finite
threshold and other to a threshold that diverges in the thermodynamic
limit. This last phase corresponds to a system that is always ordered in
the thermodynamic limit, and therefore very resilient to the effects of
external noise. These two behaviors were already observed in the CBV and
Vicsek models in non-weighted
networks~\cite{Miguel2017,Miguel.2019}. However, in the weighted case, a
new phase emerges, in which the threshold actually tends to zero in the
thermodynamic limit. This surprisingly corresponds to a system that is
always disordered for any amount of noise, however small, and indicates
a dynamics extremely susceptible to the effects of external
perturbations.

Numerical simulations in the CBV model recover the theoretical
predictions with good accuracy, with the exception of points very close
to the boundaries between regions, in which finite size effects are
stronger and larger system sizes than those considered here are
necessary. For networks of finite size, we additionally observe the
presence of a maximum in the threshold as a function of $\alpha$ for
fixed $\gamma$. This indicates that a particular weight pattern can
provide the maximum resilience against noise perturbations, by
maximizing the value of the flocking threshold. At the HMF level, this
maximum is obtained for a weight exponent $\alpha=1/2$. Simulations lead
instead to a maximum slightly depending on the degree
exponent. Furthermore, simulations of the more realistic Vicsek model
yield results that can be understood in terms of the HMF solution of the
CBV model, by simply mapping the physical limits of the noise parameters
in both models, $1$ for the Vicsek model and infinity for the CBV
model. With this mapping, simulations of the Vicsek model closely follow
the prediction and results obtained for the CBV model. We recover in
particular the presence of a region with a vanishing threshold, and
extremely susceptible to noise effects.

We finally consider the behavior of the Vicsek model in real weighted
networks representing social interactions between different animal
species. Laking a theory for real networks, we observe that the
threshold of the weighted structures is in general smaller than the one
observed in the binary (non-weighted) version of the same networks. This
indicates that the weighted pattern in real social interactions is
actually not beneficial for a flock of animals, since it reduces the
flocking threshold and thus renders the group more susceptible to
breaking in the presence of noise fluctuations. The relative difference
between the weighted and non-weighted thresholds is empirically observed
to depend on the degree of heterogeneity of the weight pattern, in a
functional form that can be approximated by a power-law. This
observation indicates that more heterogeneous patterns of weights, with
some connections much stronger than others, is again detrimental to
maintain the flock structure of the animal group.

The results presented here strengthen the equivalence between the Vicsek
and CBV model in networks~\cite{Miguel.2019} and highlight the important
effects that a social network of interactions can have on the flocking
structure of social animals. Most interestingly, they show that in some
cases the presence of a weight pattern can be counterproductive for a
flocking species, by reducing their resilience to noise or by destroying
the flocking phase altogether. The presence of such weight pattern must
thus be attributed some other adaptive benefit, that overcomes the
worsened flocking performance.

\begin{acknowledgments}
  We acknowledge financial support from the Spanish
  MCIN/AEI/10.13039/501100011033, under project No.
  PID2019-106290GB-C21. We thank Jordi Torrents for helpful comments and
  discussions.
\end{acknowledgments}

\end{document}